\documentclass[twocolumn,pre,superscriptaddress,showpacs,preprintnumbers,amsmath,amssymb,floatfix]{revtex4}
\usepackage[dvips]{graphicx}
\usepackage{amsmath}
\usepackage{bm}
\usepackage[dvips]{graphicx}
\usepackage{amsfonts}
\usepackage{bm}
\usepackage{latexsym}

\usepackage{latexsym}

\begin{document}

\title{Is there a reentrant glass in binary mixtures?}

\author{E. Zaccarelli}
\affiliation{Dipartimento di Fisica and INFM Center for Statistical Mechanics
and Complexity, Universit{\`a} di Roma La Sapienza, Piazzale Aldo Moro 2,
I-00185 Rome, Italy}
\affiliation{Institut f{\"u}r Theoretische Physik II,
Heinrich-Heine-Universit\"at D\"usseldorf,
Universit\"atsstra{\ss}e 1, D-40225 D\"usseldorf, Germany}
\author{H.\ L{\"o}wen}
\affiliation{Institut f{\"u}r Theoretische Physik II,
Heinrich-Heine-Universit\"at D\"usseldorf,
Universit\"atsstra{\ss}e 1, D-40225 D\"usseldorf, Germany}
\author{P.\ P.\ F.\ Wessels}
\affiliation{Institut f{\"u}r Theoretische Physik II,
Heinrich-Heine-Universit\"at D\"usseldorf,
Universit\"atsstra{\ss}e 1, D-40225 D\"usseldorf, Germany}
\author{F. Sciortino}
\affiliation{Dipartimento di Fisica and INFM Center for Statistical Mechanics
and Complexity, Universit{\`a} di Roma La Sapienza, Piazzale Aldo Moro 2,
I-00185 Rome, Italy}
\author{P. Tartaglia}
\affiliation{Dipartimento di Fisica and INFM Center for Statistical Mechanics
and Complexity, Universit{\`a} di Roma La Sapienza, Piazzale Aldo Moro 2,
I-00185 Rome, Italy}
\author{C.\ N.\ Likos}
\affiliation{Institut f{\"u}r Theoretische Physik II,
Heinrich-Heine-Universit\"at D\"usseldorf,
Universit\"atsstra{\ss}e 1, D-40225 D\"usseldorf, Germany}

\begin{abstract}
By employing computer simulations for a model binary mixture, we show
that a reentrant glass transition upon adding a second component
only occurs if the ratio $\alpha$ of the short-time mobilities between
the glass-forming component and the additive is sufficiently small.
For $\alpha \approx 1$, there is no reentrant glass, even if the size
asymmetry between the two components is large, in accordance with
two-component mode coupling theory.
For $\alpha \ll 1$, on the other hand, the reentrant
glass is observed and reproduced only by an effective one-component
mode coupling theory.
\end{abstract}


\maketitle

The ability to manipulate the rheological behavior of physical systems
by the addition of suitable smaller components is a task of high
technological interest. At the same time, the glassification and
melting of dynamically arrested states present the theorists with
a challenging problem.  Recent experiments \cite{Poon,Bartsch} have
revealed that the addition of non-adsorbing polymers to
sterically-stabilized colloidal dispersions induces a new glass
transition scenario. Upon adding polymers to the pure colloidal glass,
the dynamics of the colloids is first speeding up, until the glass is
molten and a fluid state is reached. By further increasing the polymer
concentration, the fluid vitrifies again and a reentrant glass
transition materializes.  This new kinetically arrested phase is
called ``attractive'' glass (in contrast to the ordinary ``repulsive''
or hard-sphere like colloidal glass), since it is generated by
effective depletion attractions between the colloids, mediated by the
polymers. The occurrence of an ``attractive'' glass seems to be
generic for any system with a short-ranged attraction and was recently
also found for colloids in a solvent of varying quality \cite{Rennie}
and in micelle-polymer mixtures \cite{Mallamace}.  The reentrant scenario
of the attractive glass had been predicted by mode-coupling theory of
the colloids by treating {\it only} the colloids explicitly, and
assuming that they interact by means of their effective,
polymer-mediated depletion interactions
\cite{Fabbian99,Bergenholtz99,Dawson00}. It has also been
observed in computer simulations of particles interacting via a
short-ranged attraction \cite{Zaccarelli,Puertas}, provided that the
range of the attraction is small enough with respect to the core
size. Nevertheless, a general understanding about the mechanisms and
the circumstances under which additives affect the glass transition 
is still lacking.  A similar situation occurs
for asymmetric molecular mixtures (such as molten salts), but just on
a different (microscopic) length scale.

In this Letter, we study systematically the question of whether and
how the dynamical properties of the additives influence the occurrence
of the glass transition, in the framework of a simple binary model
mixture.  We show thereby that the relative {\it short-time mobility}
of the constituent components plays a crucial role in determining the
occurrence of vitrification of a fluid or melting of a glass.  Our
central result is that a reentrant glass scenario requires not only a
large size asymmetry but also that the short-time mobility of the
added component be much higher than that of the glass-forming species.
In the case of a very high additive-mobility as compared to the
mobility of the glass-forming component, one can think in terms of the
adiabatic approximation: then, for an instantaneous configuration of
the glass-forming component, the additives are equilibrated,
establishing thereby the static effective interaction potential
between the former \cite{Likos_PhysRep}. It is precisely in this limit
that the concept of a {\it static}, effective, one-component
description with a depletion attraction is expected to be applicable
also in the {\it dynamical} sense.  On the other hand, if the
short-time mobilities of the two species are comparable, the depletion
potential still {\it exactly} determines the partial static structure
of the larger component \cite{Evans}, but not its dynamical
behavior. It is therefore interesting to examine whether the lack of
validity of the depletion picture in the dynamical sense has an
influence on the glass transition scenario, which is a genuinely {\it
kinetic} arrest.

From its basic formulation, the two-component mode coupling theory
(MCT) for the ideal glass transition \cite{Goetze_rev} asserts that
the latter only depends on the static partial structure factors of the
mixture and hence it is independent on the individual short-time
mobilities.  This assertion holds both for Brownian dynamics (relevant
for colloid/polymer mixtures) and for Newtonian short-time dynamics
(relevant for molecular glass formers).  Our computer simulation
studies reveal, however, that the scenario and the location of the
glass transition in the mixture depends crucially on the ratio
$\alpha$ between the short-time mobilities of the glass forming
component and of the additive.  We also show that MCT correctly
predicts the glassification scenarios for the two extreme limits: if
$\alpha \approx 1$, the {\it two-component} MCT yields no reentrance,
while for $\alpha \ll 1$, the effective {\it one-component} MCT
predicts the occurrence of a reentrant glass.  In what follows, we
first describe the model and the simulation techniques.  Our
simulation results are then compared to two-component and
one-component MCT.  Finally, estimates for the crossover towards
adiabaticity and experimental consequences are presented.

Our results are obtained for the Asakura-Oosawa (AO) binary model
\cite{AO,poon:review}, which is a simple 
prototype for a colloid-polymer mixture
but can also be used to model asymmetric non-additive hard-sphere
mixture for molecular systems. In this model, the glass-forming
particles (``colloids'') are hard spheres of diameter $\sigma_c$ and
the additives (``polymers'') do not interact with each other but
only feel the colloids as hard
spheres with an interaction radius $(\sigma_c + \sigma_p)/2$, where
$\sigma_p$ denotes the diameter of gyration of the polymer coils.
The motivation to study such an asymmetric non-additive mixture is
twofold: first, the static structure and the equilibrium phase
behavior are known, facilitating to a great extent the
understanding of the dynamical behavior. Second, the Asakura-Oosawa is
a minimal model for depletion and fluid-fluid phase separation in
mixtures, thus it contains the basic physical mechanisms driving the
occurrence of both repulsive and attractive glasses.  Since there is
no energy scale in the interactions, the thermal energy $k_BT$ scales
out and is irrelevant.  The remaining parameters for the AO-model are
the size asymmetry $q=\sigma_p/\sigma_c$, which corresponds to the
range of the attractive potential in the effective one-component
picture \cite{AO}, and the partial number densities $\rho_c$, $\rho_p$
of the colloids and polymers, which are most conveniently expressed in
terms of the partial colloid and polymer packing fractions $\eta_c =
\pi \rho_c \sigma_c^3 /6$ and $\eta_p = \pi \rho_p \sigma_p^3 /6$,
respectively.

We carried out extensive molecular dynamics (MD) simulations of the
AO-mixture, introducing the two masses $m_c$ and $m_p$ for the
colloids and the polymers. The latter determine the ratio of
short-time mobilities (or thermal velocities) $\alpha$ between
polymers and colloids as $\alpha =\sqrt{m_p/m_c}$.  MD simulations
were performed using an event-driven algorithm \cite{rapaport}.  
The relative numbers of particles and the volume of the simulation box
were chosen in order to achieve equilibrium conditions for all studied
values of $q$, $\alpha$ and $\eta_p$, within reasonable computer time
\cite{note-CPU}. Thus, we fixed the number of colloidal particles
to $N_c=500$ for $q=0.15$ and $q=0.50$, and to $N_c=40$ for $q=0.04$
\cite{size}.  Partial mean squared displacements for colloids and
polymers were calculated in order to estimate, via the Einstein
relation, the self-diffusion coefficients $D_c$ and $D_p$ of the
colloids and polymers, respectively.  To reduce statistical errors,
averages were performed over $50$ independent configurations for
$q=0.50$ and $q=0.15$, and over $10$ for $q=0.04$. The units of mass,
length, and energy are chosen as $m_c$, $\sigma_c$ and $k_BT$.

\begin{figure}
\includegraphics[width=7cm,angle=0.,clip]{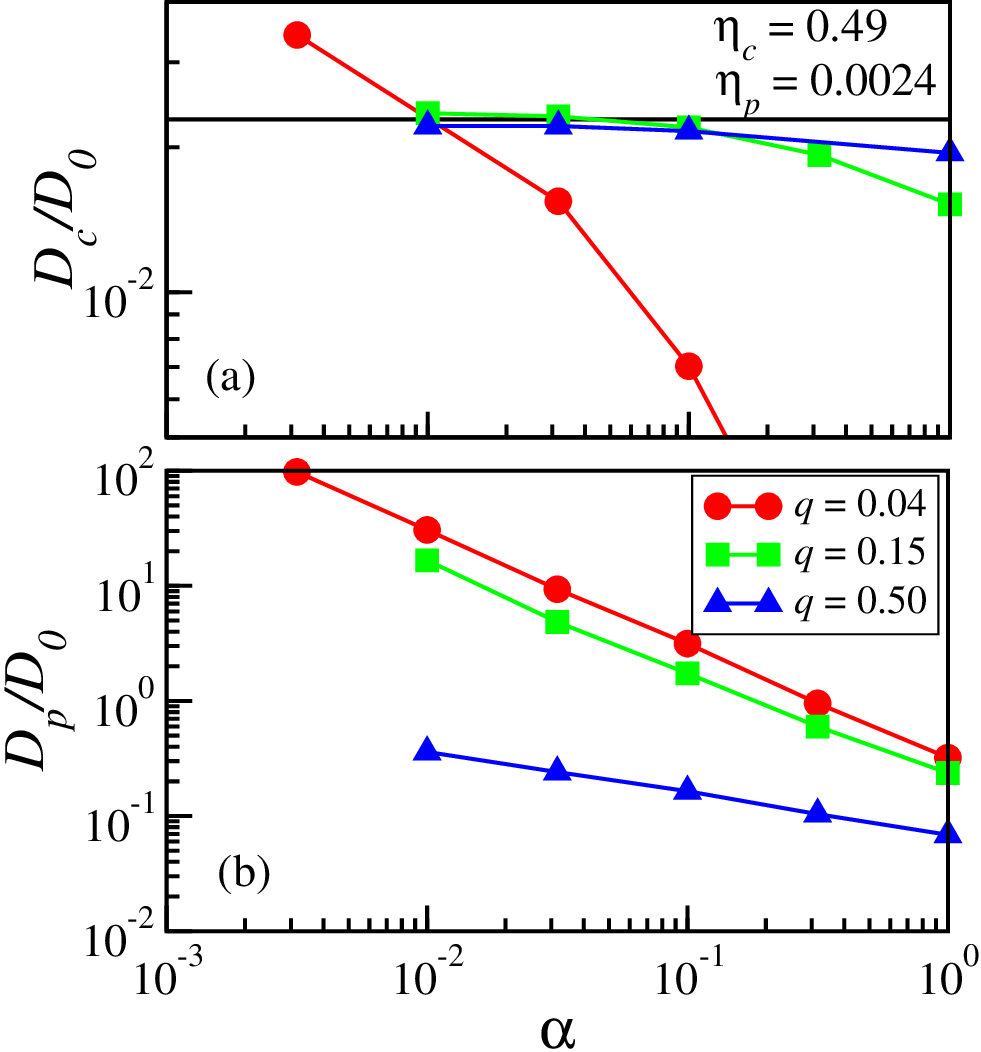}
\caption{The dependence of the diffusion coefficients on the ratio
of the short-time mobilities for (a) the colloidal particles and
(b) the polymers, at the state-point $\eta_c = 0.49$, $\eta_p = 0.0024$
and for three different $q$-values, normalized with 
$D_0 = \sqrt{k_BT\sigma_c^2/m_c}$.
The horizontal line in (a) indicates the value of $D_c$ when $\eta_p = 0$.
Note that the legends on (a) and (b) apply to both panels.}
\label{diff:fig}
\end{figure}

Simulation data for $D_c$ as a function of the mass ratio are
presented in Fig.\ \ref{diff:fig}(a) for different $q$ at a fixed
state point.  It can be seen that the diffusive motion of the colloids
depends significantly on the mass ratio and on $q$. For sufficiently
small mass ratios, the added polymer accelerates the dynamics if $q$
is small and slows it down if $q$ is large.  This behavior is in
agreement with previous studies of the dependence of the glass
transition of the colloids on the range of the effective attractive
potential \cite{Dawson00,Bergenholtz}.  However, by increasing the
mass ratio, the picture completely changes.  The added polymer {\it
always} slows down the dynamics of the colloids, irrespective of the
range of the attractive potential. This is clear evidence that,
despite of the fact that the {\it statics} is identical along each
curve at fixed $q$, the {\it dynamics} is significantly affected by
the relative mobilities of the two components.

Data for $D_p$, at the same state point and varying $q$ and $\alpha$,
are shown in Fig.\ \ref{diff:fig}(b). For small $q$, $D_p \propto
\alpha^{-1}$, which pertains to the motion of an ideal gas inside the
disordered medium formed by the colloids. For large $q$, the polymers
are caged by the colloids. This causes a significant
reduction of $D_p$ for fixed $\alpha$ and a much weaker
dependence of this quantity on $\alpha$.

\begin{figure}
\includegraphics[width=7cm,angle=0.,clip]{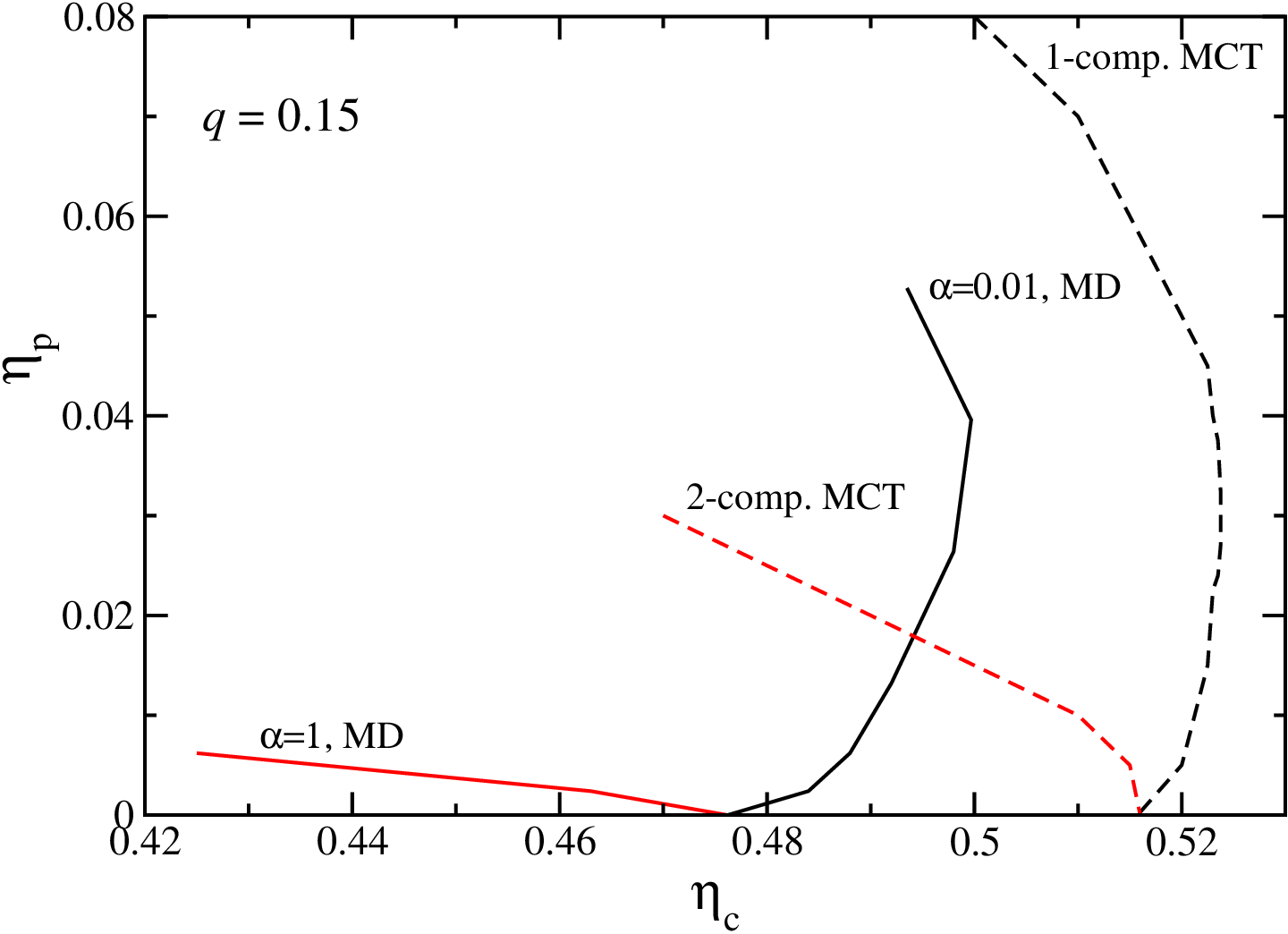}
\vspace{-0.3cm}

\caption{Isodiffusivity lines on the $\eta_c - \eta_p$-plane for
$\alpha = 0.01$ and $\alpha = 1$, accompanied by ideal glass transition
lines from the effective one-component and the two-component versions
of MCT. The results refer to $q = 0.15$.}
\label{q15:fig}
\end{figure}

\begin{figure}
\includegraphics[width=7cm,angle=0.,clip]{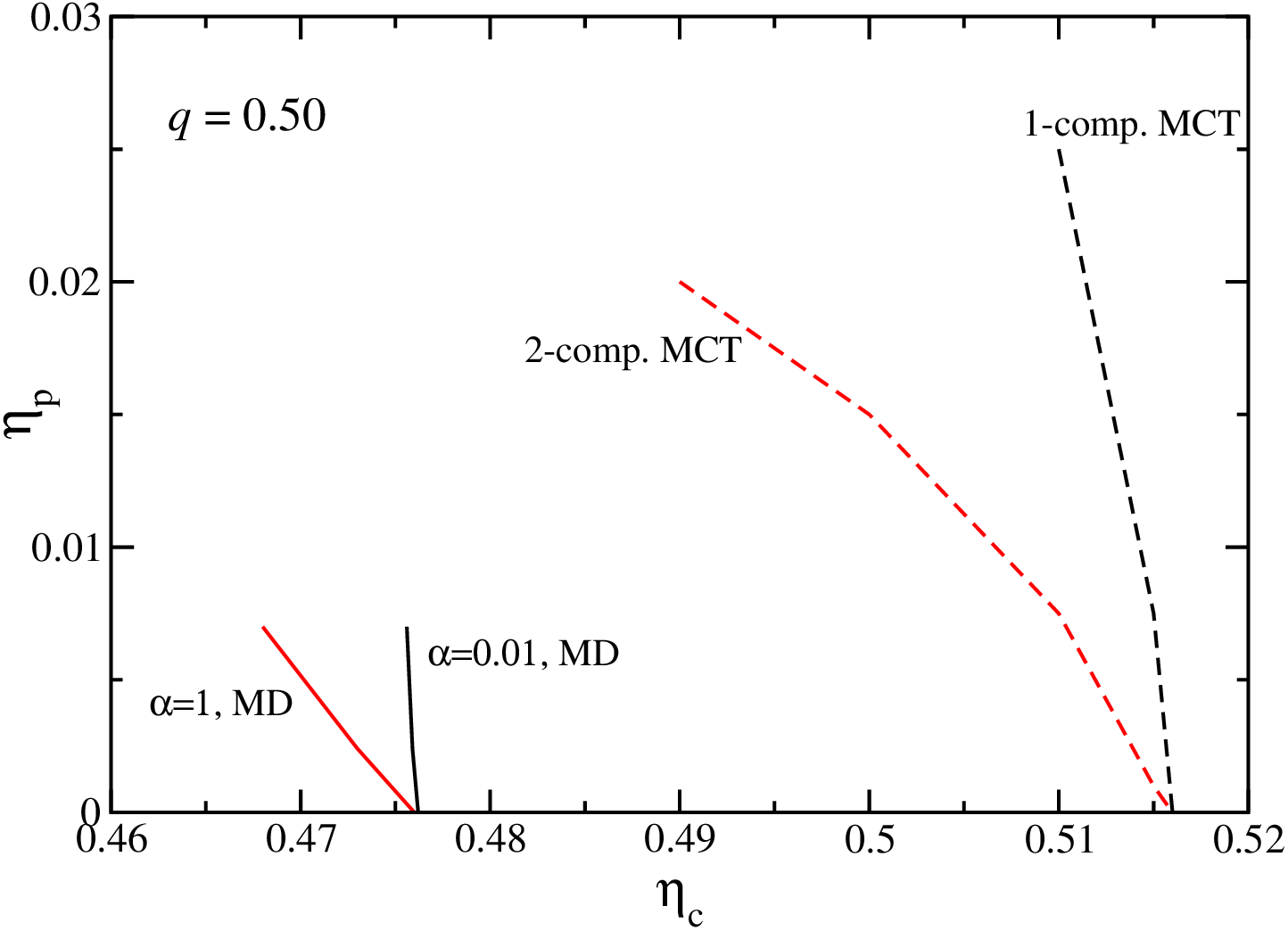}
\vspace{-0.3cm}

\caption{Same as Fig.\ \ref{q15:fig} but for $q = 0.50$.}
\label{q50:fig}
\end{figure}

In order to obtain a broader picture of the dynamical behavior of the
system on the entire $\eta_c - \eta_p$-plane, we estimate the
isodiffusivity lines from the MD simulations. These can be considered
as precursors of the incipient glass transition line and provide an
estimate of the shape of the latter \cite{foffi}. Results for the
isodiffusivity lines for $q = 0.15$ are shown in Fig.\ \ref{q15:fig}
and for $q = 0.50$ in Fig.\ \ref{q50:fig}. In each
Figure, results for two very different mass ratios, corresponding to
$\alpha=1$ and $\alpha=0.01$, are shown. For $q = 0.15$, two different
kinds of behavior result: for $\alpha=0.01$, a clear reentrance is
observed, while for $\alpha=1$, the addition of polymers always slows
down the dynamics. For a smaller size asymmetry, $q=0.50$, there is
always polymer-induced vitrification and no reentrance, but the impact
of added polymer on the glass formation, embodied in the slope of the
isodiffusivity curves on the $\eta_c-\eta_p$-plane, depends strongly
on $\alpha$. For small mobility ratios, the isodiffusivity curves are
almost vertical, pointing to an insensitivity of the diffusion
coefficient on the amount of added polymer. For larger $\alpha$,
addition of polymer considerably slows down the dynamics.

In order to provide a criterion for the applicability of MCT to
different regimes of mobility ratios, we have calculated the location
of the ideal glass transition within the framework of this
theory. 
We have solved the implicit
equations for the matrix of partial non-ergodicity parameters ${\bf
F}(q)$,
\begin{equation}
{\bf F}(q)={\bf S}(q)-\{{\bf S}^{-1}(q)+{\cal F}[{\bf F}(q)]\}^{-1}
\label{eq:mct}
\end{equation}
where ${\bf S}(q)$ is the matrix of partial static structure factors,
and ${\cal F}[{\bf F}(q)]$ is the long-time limit memory kernel within
MCT approximation \cite{Goetze_rev}. More specifically, it is a
quadratic functional of ${\bf F}(q)$, coupling all different
wave-vectors, by means of coupling coefficients that only depend on
partial static structure factors and partial concentrations of the
system. A glass transition occurs when, upon changing the control
parameters, the solution of Eq.\ (\ref{eq:mct}) jumps from zero
(ergodic state) to a finite value (glassy state). Note that, within
MCT, inertia parameters drop out, hence the location of the ideal
glass line is independent of $\alpha$. Eq.\ (\ref{eq:mct}) holds also
for the one-component case, where ${\bf S}(q)$ and ${\bf F}(q)$ are $1
\times 1$ matrices.

The partial structure factors needed as input for the two-component
and one-component mode coupling theory were obtained from the recently
developed fundamental measure density functional theory \cite{schmidt00}.
This theory yields analytical
expressions for the partial structure factors $S_{ij}(k)$ with
$i,j=c,p$ which compare well with simulation data. In the case of
the effective one-component MCT, ${\bf S}(q) = S_{cc}(q)$.

Results from both effective one- and two-component MCT are shown in
Figs.\ \ref{q15:fig} and \ref{q50:fig} for $q=0.15$
and $q=0.50$, respectively.  In the first case, the two-component MCT
predicts a vitrification of the fluid upon addition of polymers, while
the effective one-component MCT predicts a reentrant scenario. In the
second case, both the two component and the effective one-component
MCT predict a vitrification of the fluid upon addition of polymers.
In this last case, the slope of the ideal glass line is almost
vertical for the one-component calculation.

Since the isodiffusivity lines are precursors of the shape in the
$\eta_c - \eta_p$ plane of the glass transition line, 
the comparison between these lines and the MCT calculations
allows us to conclude that the MD results for small $\alpha$ are
properly captured by the effective one-component MCT, while the MD
results for large $\alpha$ are captured by the full two-component MCT
calculation.

The ratio $\alpha$ needed to reach the adiabatic limit-regime, in
which an effective one-component description applies dynamically can
be estimated as follows.  During the time necessary for a colloidal
particle to move along a distance $\sigma_p$, there has to be a
sufficiently large number of collisions, exceeding at least some
typical threshold value $N_0 \approx 100$.  This is necessary in order
to provide enough statistics to feel the effective interaction
dynamically \cite{gerrit}. The resulting adiabaticity condition reads
as
\begin{equation}
\alpha \lesssim 12 \eta_p/(q^2 N_0). 
\label{criterion:eq}
\end{equation} 
For the two chosen two cases, $\alpha =0.01$ and $\alpha =1$ in Figs.\
\ref{q15:fig} and \ref{q50:fig}, we realize that, for a typical
$\eta_p$, the abiabaticity criterion (\ref{criterion:eq}) is fulfilled
in the former but not in the latter case. This in turn explains why
the effective depletion potential is dynamically meaningful for the
case $\alpha =0.01$, so that the effective one-component MCT is
applicable.  On the other hand, for $\alpha =1$ adiabaticity is not
achieved and both the colloids and polymers are cage formers, so that
the two-component MCT has to be employed in the study of the
vitrification transition.

The same scaling arguments employed for the derivation of criterion
(\ref{criterion:eq}) can be carried out for short-time {\it Brownian}
dynamics, for which $\alpha = D_c^{(0)}/D_p^{(0)}$, with $D_c^{(0)}$
and $D_p^{(0)}$ denoting the short-time diffusion coefficients of the
colloids and the polymers, respectively, in their common solvent. In a
mixture of uncharged colloids and polymers, these coefficients scale
with the inverse radii of the particles, according to Stokes'
expression, hence they are coupled to $q$ via $D_c^{(0)}/D_p^{(0)}=q$.
Then the critical asymmetry $q$ below which adiabaticity holds is
given by $q_c = (12\eta_p/N_0)^{1/3}$, which is close to $0.1$ for
typical values of $\eta_p$ and $N_0$, On the other hand, in mixtures
of {\it charged} suspensions, the {\it physical} hard core diameters
$\sigma_c^{(0)}$ and $\sigma_p^{(0)}$ are different and independent
from the effective interaction diameters that enter in the $q$
ratio. Indeed, effective interaction diameters $\sigma_c$ and
$\sigma_p$ are dictated by the long-range Coulomb interactions, thus
they can greatly exceed the physical ones.  The charge on the
particles provides in this case the physical parameter that allows
tuning of the physically relevant size ratio $q$.  In this case, the
adiabaticity criterion (\ref{criterion:eq}) reads as
$\sigma_p^{(0)}/\sigma_c^{(0)} \lesssim 12 \eta_p/(q^2 N_0)$.

In concluding, we have shown that a sufficient asymmetry in the
short-time mobilities is necessary for the reentrant glass-scenario to
materialize in highly asymmetric binary mixtures.  For glass-forming
mixtures governed by molecular dynamics, the glass transition depends
on the mass ratio between the two components \cite{footnote}, but huge
mass asymmetries are needed to achieve the adiabatic limit. Vastly
different mass ratios can be realized in binary mixtures of dusty
plasmas \cite{thomas}, whose dynamics is almost Newtonian.  On the
basis of the considerations put forward in this Letter, the size
asymmetry $q$ for colloid-polymer mixtures has to be smaller than
$0.1$ in order to get a reentrant glass.  This condition is stronger
than the one obtained by using a priori an effective one-component
picture. Indeed, in the present calculation one-component MCT would
predict a reentrant glass for $q\lesssim 0.3$, while for previous
studies of effective one-component pictures
\cite{Dawson00,Bergenholtz} $q$ up to at least $0.2$ was found to have
an attractive glass \cite{q-note}.  This finding helps to explain why
in experimental studies \cite{Poon,Mallamace}, a very small size
asymmetry was indeed necessary in order to obtain an attractive glass.
The most striking consequences are obtained for a mixture of charged
colloids: a reentrant glass can be lost if the physical core of the
high-charge particles is decreased, even if the charges are kept
constant. By tuning the short-time mobility, the glass formation upon
addition of a second repulsive component can be tailored, opening a
way for external manipulation of the rheology of the mixture.

If the kinetic glass transition in a mixture is calculated within mode
coupling theory, it is not known a priori whether the two-component or
the effective one-component versions of the theory has to be employed.
We have shown that it is the adiabaticity criterion
(\ref{criterion:eq}) that determines which theory must be applied. The
impact of the $\alpha$-dependence of the glass transition should also
have unexplored important consequences in polymer mixtures, such as,
e.g., mixtures of star polymers of different arm numbers and radii, in
which the lighter polymeric component has been used as a modifying
agent of the flow-behavior of the heavier one
\cite{Vlassopoulos1,Vlassopoulos2}.

We are grateful to S. Buldyrev for the MD code, and to W. G{\"o}tze,
F. Scheffold and G. Vliegenthart for helpful discussions. Financial support
within the Deutsche Forschungsgemeinschaft (SFB TR6), the EU
Marie-Curie Network, MIUR COFIN 2002 and FIRB is gratefully
acknowledged.


\begin{thebibliography}{10}
\bibitem{note-CPU} Note that for $q=0.04$ at the smallest $\alpha$,
the results for the state point reported in Fig.\
\protect\ref{diff:fig}, where $N_p=3960$, required 50 CPU days on a
fast AMD Athlon processor.
\bibitem{size} Size-dependent effects were accounted for by rescaling the
colloid diffusivity values to the ones for $N_C=500$.
\bibitem{q-note} Note that between the AO model and the square well
model used in \protect\cite{Dawson00} a factor of $2$ in $q$ must be
taken into account, due to the different potential shapes.
\end{thebibliography}
\end{document}